\documentclass[pra,notitlepage,superscriptaddress,nofootinbibfloatfix,amsmath,amsfonts,amssymb,longbibliography]{revtex4-2}%
\usepackage{braket}
\usepackage{subcaption}
\usepackage{tabularx}
\usepackage{dsfont}
\usepackage{amsmath,amsfonts,amssymb}
\usepackage{float}
\usepackage{graphicx}% Include figure files
\usepackage{dcolumn}% Align table columns on decimal point
\usepackage{bm}% bold math
\usepackage{hyperref}% add hypertext capabilities

\usepackage{amsmath,amsfonts,amssymb,color}
\usepackage{amsthm}
\usepackage{leftidx}
\usepackage{graphicx}
\usepackage{comment}
\usepackage{xcolor}
\usepackage{dcolumn}
\usepackage{bm}
\usepackage{epstopdf}
\usepackage{epsfig}
\usepackage{environ}
\usepackage{pdfcomment}

\usepackage{float}
\usepackage[T1]{fontenc}
\usepackage[latin9]{inputenc}
\usepackage{setspace}
\usepackage{esint}

\begin{document}

%\preprint{APS/123-QED}

\title{{On the quantization of AB phase in nonlinear systems}}% Force line breaks with \\

%\author{Xi Liu}
% \altaffiliation[Also at ]{Physics Department, XYZ University.}
% \author{Senmao Tan}
%\author{Longwen Zhou}%
%\affiliation{%


%\author{Qing-hai Wang}
%\affiliation{
%\affiliation{
%\author{Jiangbin Gong}
%\affiliation{%


\author{Xi Liu}
%\email{xxx@u.nus.edu}
\affiliation{NUS Graduate School - Integrative Sciences and Engineering Programme (ISEP), National University of Singapore, Singapore 119077, Singapore}

\author{Qing-hai Wang}
%\email{qhwang@nus.edu.sg}
\affiliation{Department of Physics, National University of Singapore, Singapore 117551, Singapore}

\author{Jiangbin Gong}
\email{phygj@nus.edu.sg}
%\affiliation{NUS Graduate School - Integrative Sciences and Engineering Programme (ISEP), National University of Singapore, Singapore 119077, Singapore}
\affiliation{Department of Physics, National University of Singapore, Singapore 117551, Singapore}
\affiliation{Centre for Quantum Technologies, National University of Singapore, Singapore 117543, Singapore}
\date{\today}% It is always \today, today,

\begin{abstract}
 {Self-intersecting energy band structures in momentum space can be induced by nonlinearity at the mean-field level, with the so-called nonlinear Dirac cones as one intriguing consequence. Using the Qi-Wu-Zhang model plus power law nonlinearity, we systematically study in this paper the Aharonov-Bohm (AB) phase associated with an adiabatic process in the momentum space, with two adiabatic paths circling around one nonlinear Dirac cone.  Interestingly, for and only for Kerr nonlinearity, the AB phase experiences a jump of $\pi$ at the critical nonlinearity at which the Dirac cone appears or disappears, whereas for all other powers of nonlinearity the AB phase always changes continuously with the nonlinear strength. Our results may be useful for experimental measurement of power-law nonlinearity and shall motivate further fundamental interest in aspects of geometric phase and adiabatic following in nonlinear systems.    }
\end{abstract}

%Use showkeys class option if keyword
\maketitle

\section{Introduction}
\label{sec:introduction}
The dynamics depicted by a nonlinear discretized Sch\"{o}dinger equation (NDSE) can be extremely rich, including the emergence of many-dimensional chaos, solitons, and breathers etc. The problem can be much reduced by assuming the translational invariance of a wave under consideration. With this assumption the main physics is about the features of Bloch waves,  the associated energy bands, and how they respond to changes in the parameters of a nonlinear system.    Interestingly, the nonlinear Bloch bands of NDSE can induce gapless band structures absent in linear systems, such as 2-dimensional (2D) noninear Dirac cones \cite{nonlinearDiracCones} induced by Kerr nonlinearity~\cite{new_2011}.   Even more peculiar,  such nonlinear Dirac cones  are formed by exotic nonlinear energy bands in a subregime of the Brillioun zone~\cite{nonlinearLandauZener,nonlinearDiracCones,nonlinearityInducedTopologicalPhysics,LiuJPhysRevA.66.023404,WitthautPhysRevA.73.063609,ZhangQ,Zhang_2008}.  

As a close analog to a setting in real space to measure the Aharonov-Bohm (AB) phase around a singularity point with magnetic flux,  let us now imagine two adiabatic paths, in the momentum space, circling around a band-crossing point.  If we adiabatically change the Bloch momentum, so as to guide the Bloch wave to evolve along the two adiabatic paths, the final phase difference thus generated between the two adiabatic paths is termed the nonlinear AB phase \cite{nonlinearDiracCones}. One may na\"{i}vely think of the following:  provided  that the dynamical phases between the two adiabatic paths are identical and hence have zero contribution to the phase difference of interest, the obtained AB phase would be just the Berry phase associated with the band degeneracy point.  The actual physics turns out to be more interesting than just a Berry phase. Because of nonlinearity, any small deviation of the adiabatically following state from the instantaneous Bloch wave causes a tiny correction to the dynamical phase, and accumulation of such tiny corrections over the entire adiabatic protocol yields an unfamiliar geometrical phase on top of the expected Berry phase.  Remarkably, as a possible means of topological charaterization of nonlinear Dirac cones, it is found in Ref.~\cite{nonlinearDiracCones} that the nonlinear AB phase around nonlinear Dirac cones induced by Kerr nonlinearity added to the so-called Qi-Wu-Zhang (QWZ) model~\cite{QWZPhysRevB.74.085308}
is quantized in $\pi$, whereas the Berry phase is not quantized (thus in sharp contrast to a variery of linear systems, where the Berry phase around a Dirac cone is quantized in $\pi$~\cite{Novoselov2005,Zhang2005,Ando1998,PhysRevLett.82.2147}).
Echoing with the finding in ~\cite{nonlinearDiracCones}, Ref.~\cite{nonlinearityInducedTopologicalPhysics} found $\pi$-quantization of a nonlinear Zak phase and Ref.~\cite{nonlinearWyleSemimetals} further confirmed the $\pi$-quantization of the nonlinear AB phase around a nodal line induced by Kerr nonlinearity.

The goal of this work is entirely focused on aspects of the nonlinear AB phase around Dirac cones induced by general power law nonlinearity ~\cite{MilovanovPhysRevE.103.052218,WAZWAZ2006802,KILIC201764,Sulem2004TheNS,OSMAN2019102157,MIRZAZADEH2017178,DAI2017239,MIRZAZADEH20144246,Biswas2006IntroductionTN}. In this way, it becomes possible to answer whether the previously obtained AB phase quantization is unique to Kerr nonlinearity and if so, why there is such uniqueness. Using the QWZ model~\cite{QWZPhysRevB.74.085308} as the linear limit, we are able to analytically show that Kerr nonlinearity happens to be a critical case among all kinds of power law nonlinearity. Specifically, for any nonlinearity other than the cubic order, the $\pi$-quantization of nonlinear AB phase does not exist.  Our analytical results are further confirmed by direct numerical simulations.

\section{Hamiltonian and energy spectrum}
The momentum-space Hamiltonian is composed of a QWZ model with power law nonlinearity characterized by a parameter $p$:
\begin{align}
   \widehat{H}(\psi)&=J_{1}\sin k_{1}\sigma_{1}+J_{2}\sin k_{2}\sigma_{2}+\beta(k_{1},k_{2})\sigma_{3}+g\begin{bmatrix}
        |\psi_{1}|^{2p} &0\\
        0 & |\psi_{2}|^{2p}
    \end{bmatrix},
    \label{eqnHamiltonian}
\end{align}
where $\sigma_i$ are Pauli matrices and $\psi_a$ are two components of the wavefunction, $\psi=\begin{bmatrix}
        \psi_{1}\\
        \psi_{2}
    \end{bmatrix}$.
The normalization of the wavefunction means that $|\psi_1|^2+|\psi_2|^2=1$. The nonlinearity parameter $p$ is a nonnegative real number. The Kerr nonlinearity corresponds to $p=1$. The parameters $k_{1}$ and $k_{2}$ are two quasimomenta, whose values will be adiabatically tuned in order to implement an actual adiabatic protocol to generate the nonlinear AB phase. 

To solve the nonliner eigenvalue problem,
\begin{align}
    \widehat{H}(\psi)\ket{\psi}&=E\ket{\psi}
    \label{eqnEigOriginal},
\end{align}
we introduce a real parameter $x$ as
\begin{align}
    \psi_{1} &=\sqrt{\frac{1+x}{2}}, \qquad
    \psi_{2} =\sqrt{\frac{1-x}{2}}e^{i\varphi}. \label{eqn:y20}
\end{align}
We will see later that the angular variable $\varphi$ is the same as in the Fig.~\ref{figDynPath}.
%\begin{align}
%   x:=|\psi_{1}|^{2}-|\psi_{2}|^{2}. 
%   \label{eqn:x}
%\end{align}
It turns out that $x$ is the central quantity for expressing energy, dynamical phase, Berry phase, and nonlinear AB phase. It can be shown that the instantaneous eigenenergy is
\begin{align}
    E&=\frac{\beta}{x}+\frac{g}{x}\left[\left(\frac{1+x}{2}\right)^{p+1}-\left(\frac{1-x}{2}\right)^{p+1}\right], \label{eqnEInxFull}
\end{align}
where $x$ satisfies the following algebraic equation, 
\begin{align}
    \frac{1-x^2}{x^2}\left\{\beta+\frac{g}{2}\left[\left(\frac{1+x}{2}\right)^p-\left(\frac{1-x}{2}\right)^p\right]\right\}^{2} &= |\gamma|^2,
    \label{eqnFullf}
\end{align}
with $\gamma:=J_{1}\sin k_{1}-iJ_{2}\sin k_{2}$.
%By solving the equation $f(x)=0$ for different values of $k_{1}$ and $k_{2}$, we can get the values of energy spectrum using Eq.~\ref{eqnEInxFull}. 

\begin{figure}
    \centering
    \includegraphics[width=1\textwidth]{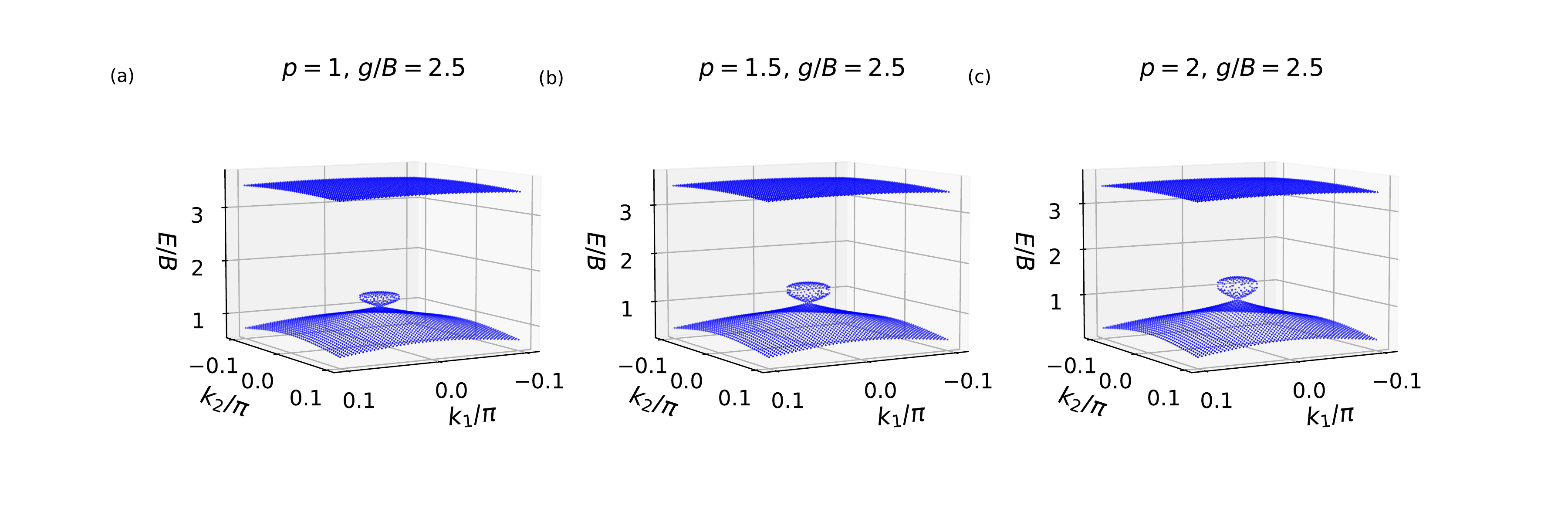}
    \caption{Nonlinear band structure for small momenta in the vicinity of the origin, i.e., for small values of  $|k_1|$ and $|k_{2}|$, with the power law nonlinearity parameter $p=1.5,2,2.5$ and the nonlinear strength parameter $g=2.5B$. See the main text for details of the system parameters.  The Dirac cone emerges from the lower energy band.}
    \label{fig3Surfaces}
\end{figure}

In order to have a Dirac point in the energy spectrum, the energy must be doubly degenerate at $k_{1}=k_{2}=0$. Since $\gamma=0$ at the this point, $x$ must satisfy
\begin{align}
    \beta(0,0)+ \frac{g}{2}\left[\left(\frac{1+x}{2}\right)^p-\left(\frac{1-x}{2}\right)^p\right]&=0.
    \label{eqngx}
\end{align}
For simplicity, we choose 
\begin{align}
    J_{1}&=J_{2}:=B,\\
    \beta(k_1,k_2)&=B(-1+\cos k_{1}+\cos k_{2}).
\end{align}
Hence $\beta(0,0)=B$. It is clear that the nonlinearity strength $g$ and energy $E$ can be scaled in terms of $B$. Energy spectra with $p=1,1.5,2$ and $g=2.5B$ are shown in Fig.~\ref{fig3Surfaces}, where Dirac cone is clearly visible around the origin. A perturbative analysis of energy spectrum near the Dirac cone can be found in Appendix~\ref{appEig}. 

\begin{comment}
\begin{figure*}
   \centering
\begin{subfigure}{1\columnwidth}
\centering
\includegraphics[width=0.45\textwidth]{p1.pdf}
%\end{subfigure}\hfill
%\begin{subfigure}{1\columnwidth}
\centering
\includegraphics[width=0.45\textwidth]{p1.5.pdf}
\end{subfigure}
\medskip

\begin{subfigure}{1\columnwidth}
\centering
\includegraphics[width=0.45\textwidth]{p2.pdf}
\end{subfigure}

    \caption{Nonlinear band structure for small momenta in the vicinity of the origin, i.e., for small values of  $|k_1|$ and $|k_{2}|$, with the power law nonlinearity parameter $p=1.5,2,2.5$ and the nonlinear strength parameter $g=2.5B$. See the main text for details of the system parameters.  The Dirac cone emerges from the lower energy band.}
    \label{fig3Surfaces}
\end{figure*}
\end{comment}

\section{Dynamics of Adiabatic Following}
\begin{figure*}
    \centering
    \includegraphics[width=0.6\textwidth]{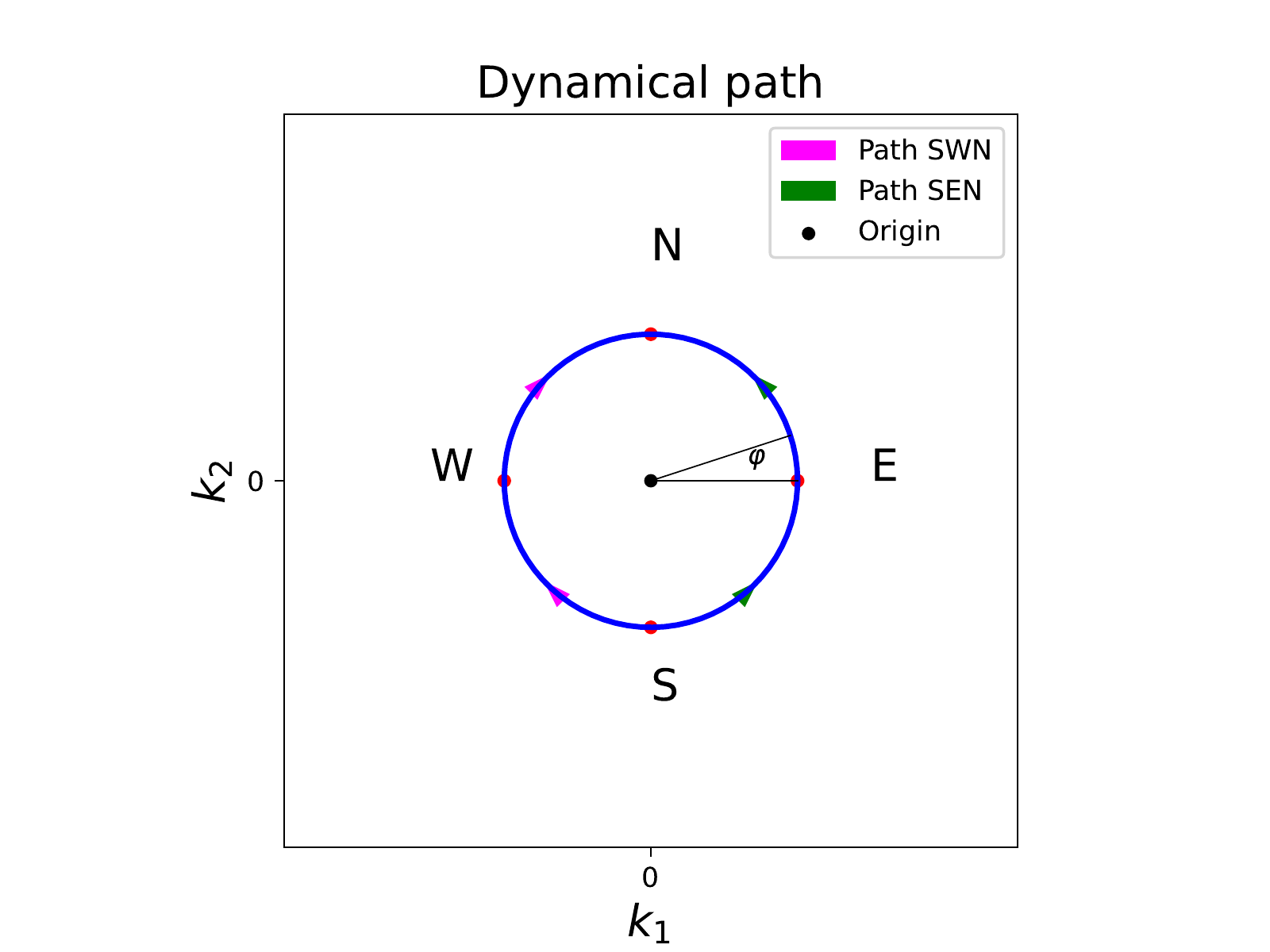}
    \caption{Dynamical paths in the momentum space. Path SWN and path SEN are symmetric halves of the perimeter of the circle. The system starts its adiabatic following at point S, and ends at point N. The two paths are parameterized by $\varphi$ in the main text.}
    \label{figDynPath}
\end{figure*}

To obtain the nonlinear AB phase, let us consider two adiabatic paths along a small circle around the origin $k_{1}=k_{2}=0$.  As shown in Fig.~\ref{figDynPath}, starting at the same point S, along each path the system is guided to move along one half of the perimeter of the circle using the same amount of time. The two adiabatic paths are ``recombined" at the end of the evolution at point N. As introduced in Sec.~\ref{sec:introduction}, the phase difference acquired by the system between two adiabatic paths is called the nonlinear AB phase. Clearly, the nonlinear AB phase here is the sum of the dynamical phase difference and the Berry phase associated with the closed loop around the band-degeneracy point.   We shall study below  the possible AB phase quantization for a varying nonlinearity strength $g$ and for different nonlinear parameters $p$. The quasimomenta $k_{1}$ and $k_{2}$ associated with two spatial dimensions are parameterized by $\varphi$ and will be made to adiabatically change. 

At the starting point S, the system is assumed to be prepared in the Bloch eigenstate at momentum space location S.  As the system adiabatically evolves along the path SEN or SWN, the time-evolving state deviates from the instantaneous eigenstate along the path, with the tiny deviation  at the order of the adiabatic parameter $\varepsilon$. The slower the rate of adiabatic change is, the smaller $\varepsilon$ is, and the less the deviation.  Here nonlinearity plays a key role. That is, the dynamical phase also obtains a correction at the order of $\varepsilon$. Since the  total evolution time is of order $O(\varepsilon^{-1})$, the $O(\varepsilon)$ term in this phase correction will contribute an $\varepsilon$-independent term through accumulation, yielding a geometric phase term out of the dynamical phase.  This will not occur in linear terms because such correction accumulated over the entire adiabatic process is at most of the order of $\varepsilon$, which vanishes for sufficiently slow adiabatic protocols. 

The dynamics of the states is governed by the time-dependent Schr\"{o}dinger equation,
\begin{align}
    i\ket{\dot{\Psi}}&=\widehat{H}(\Psi)\ket{\Psi},
    \label{eqnEvolution}
\end{align}
where the Hamiltonian is given by Eq.~(\ref{eqnHamiltonian}) with $\psi$ being replaced by $\Psi$. Here the overhead dot denotes the time derivative. We will solve this equation up to the order of $\varepsilon$ as described above. Through the lengthy computation as illustrated in Appendix \ref{appDynamics}, we obtain the instantaneous change rate of the overall phase of a time-evolving state as
\begin{align}
    \dot{\theta} & \sim -E  -\frac{1-x}{2} \dot{\varphi} + gp \frac{x(1-x^2)}{4\Delta} \left[\left(\frac{1+x}{2}\right)^p - \left(\frac{1-x}{2}\right)^p\right] \dot{\varphi},
    \label{eqn:dynphase}
\end{align}
with 
\begin{align}
    \Delta &:= \beta+\frac{g}{2}\left[(1-px+px^2)\left(\frac{1+x}{2}\right)^p - (1+px +px^2)\left(\frac{1-x}{2}\right)^p\right].
    \label{eqn:Delta}
\end{align}
We recognize that the circular integration of the second term in Eq.~(\ref{eqn:dynphase}) is nothing but the Berry phase $\theta_B$, because it assumes the same form as in the linear limit.  The rest of the phase is from the dynamical phase $\theta_D$, which contains two parts: the first part comes from the instantaneous eiegnenergy $E$ and the second part from the third term in Eq.~(\ref{eqn:dynphase}) as a new contribution from the nonlinearity. Specifically, 
\begin{align}
    \theta_B &:= - \oint \frac{1-x}{2} d\varphi,\\
    \theta_D &:= -\int E dt 
    %\nonumber\\    {}&
    + g p \int \frac{x(1-x^2)}{4\Delta} \left[\left(\frac{1+x}{2}\right)^p - \left(\frac{1-x}{2}\right)^p\right] d\varphi. \label{eqn:thetaD}
\end{align}
In the event that the Dirac cone does exist at the point $k_{1}=k_{2}=0$, the obtained phase difference between the two adiabatic paths described in Fig.~\ref{figDynPath} then becomes the nonlinear AB phase $\theta_{AB}$. Since the two adiabatic paths are symmetric by construction and that they take the same amount of time, the leading term in Eq.~(\ref{eqn:thetaD}) contributes the same in each of the two paths. Thus, the difference of the dynamical phases between two paths comes from the second term of Eq.~(\ref{eqn:thetaD}) only. Thus, the total nonlinear AB phase is
\begin{align}
    \theta_{AB} & := \theta_B + \delta\theta_D \nonumber\\
    &\sim - \pi (1-x)  
    %\nonumber\\     {}&
    + \pi g p \frac{x(1-x^2)}{2\Delta} \left[\left(\frac{1+x}{2}\right)^p - \left(\frac{1-x}{2}\right)^p\right] .
\end{align}

Note that we take into account that the paths are chosen to be close to the Dirac cone (so that the cones indeed have linear dispersion relations), namely, $|k_1|$ and $|k_2|$ are small at all times.  The leading behavior of the dynamical phase difference term is then found to be
\begin{align}
    \delta \theta_{D}&\sim \pi g p \frac{x_0(1-x_0^2)}{2\Delta_0} \left[\left(\frac{1+x_0}{2}\right)^p - \left(\frac{1-x_0}{2}\right)^p\right], \label{eqnThetaDLeading}
\end{align}
where $\Delta_0$ is $\Delta$ evaluated at $x=x_0$ and $k_1=k_2=0$, $x_{0}$ is the solution of Eq.~(\ref{eqngx}), and
\begin{align}
    \Delta_0 & = - \frac{gp x_0}{2} \left[(1-x_0) \left(\frac{1+x_0}{2}\right)^p + (1+x_0)\left(\frac{1-x_0}{2}\right)^p  \right] .\label{eqnD1Leading}
\end{align}
For the Berry phase, the leading behavior is
\begin{align}
    \theta_{B}&\sim -\pi (1-x) \sim-\pi (1- x_{0}).\label{eqnThetaBLeading}
\end{align}

%Similarly, the leading behavior of Berry phase difference of the two beams is given by 
%\begin{align}
%\theta_{B}&\sim i\oint \braket{y_{0}|\partial_{\phi}y_{0}}=-\pi (1-x).\label{eqnThetaB}
%\end{align}
%Since the paths are very close to the origin, we may express the dynamical phase around the origin $\theta_{D}$ and Berry phase around the origin $\theta_{B}$ using the leading behavior as $x\sim x_{0}$. In terms of $x_{0}$, these quantities are given by 

%Eq.~(\ref{eqnThetaBLeading}), Eq.~(\ref{eqnThetaDLeading}) and Eq.~(\ref{eqnD1Leading}). 
\begin{figure*}[hbtp!]
\centering
\begin{subfigure}{1\columnwidth}
\centering
\includegraphics[width=0.45\textwidth]{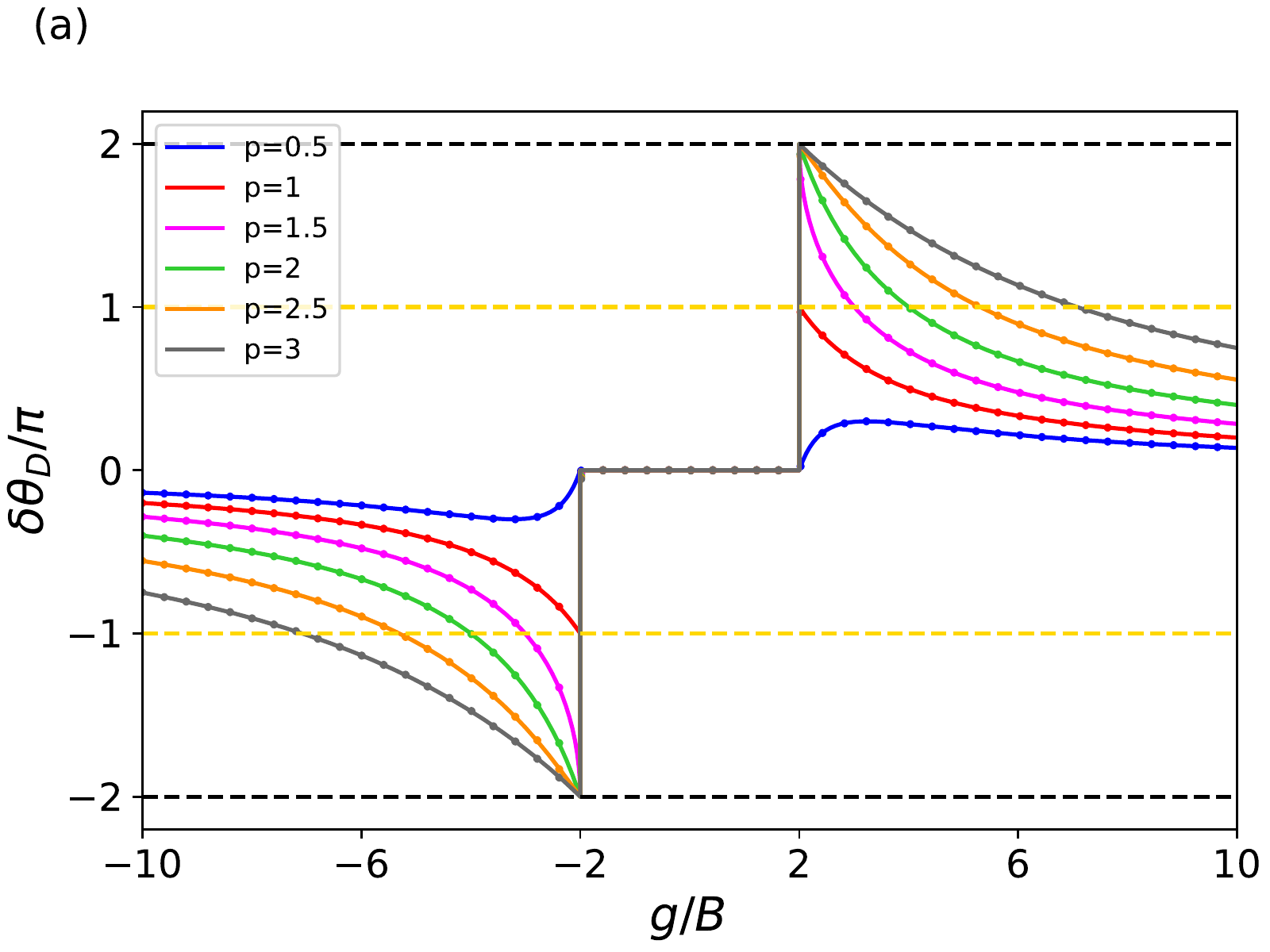}
%\end{subfigure}\hfill
%\begin{subfigure}{1\columnwidth}
\centering
\includegraphics[width=0.45\textwidth]{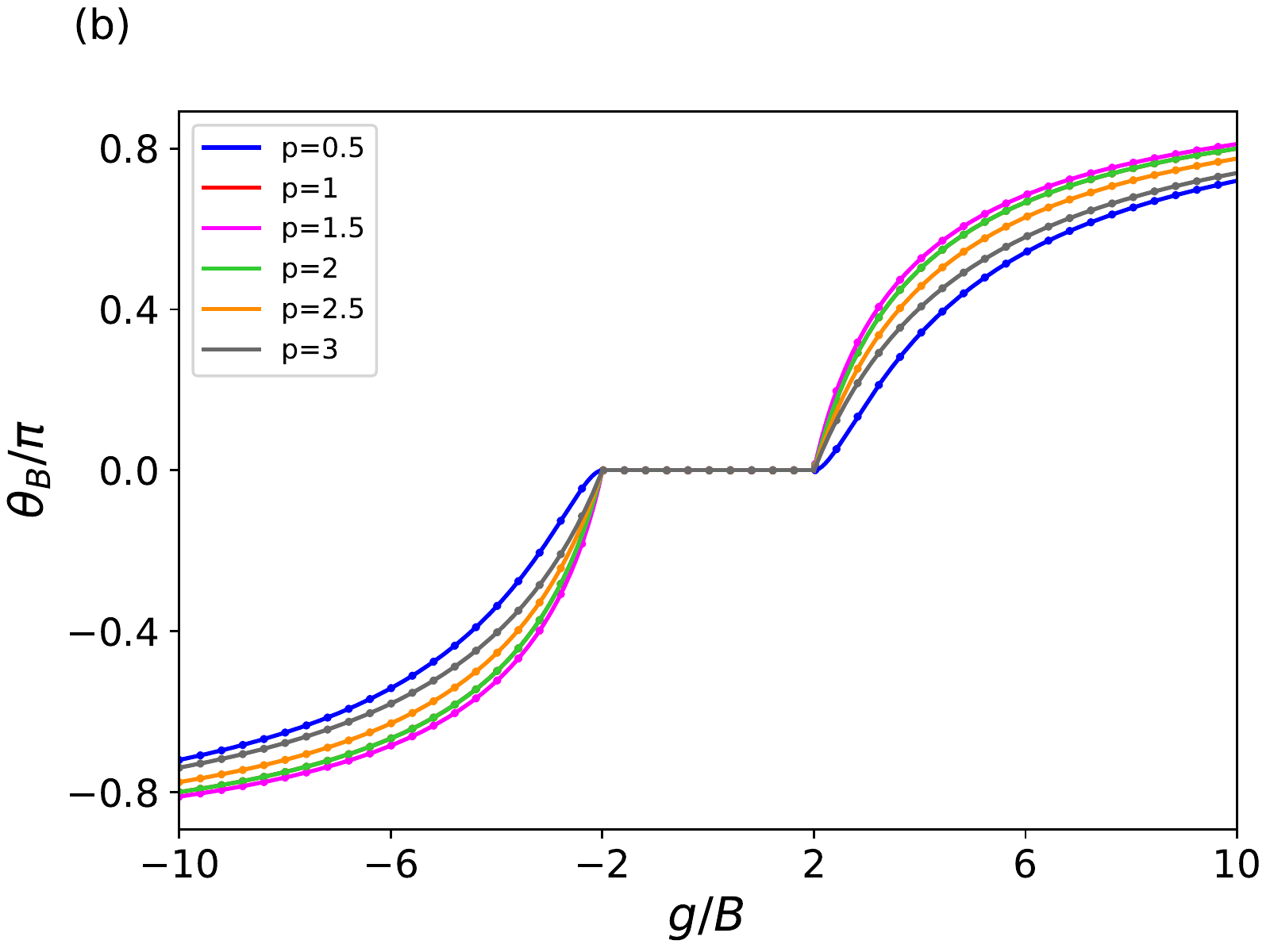}
\end{subfigure}
\medskip

\begin{subfigure}{1\columnwidth}
\centering
\includegraphics[width=0.45\textwidth]{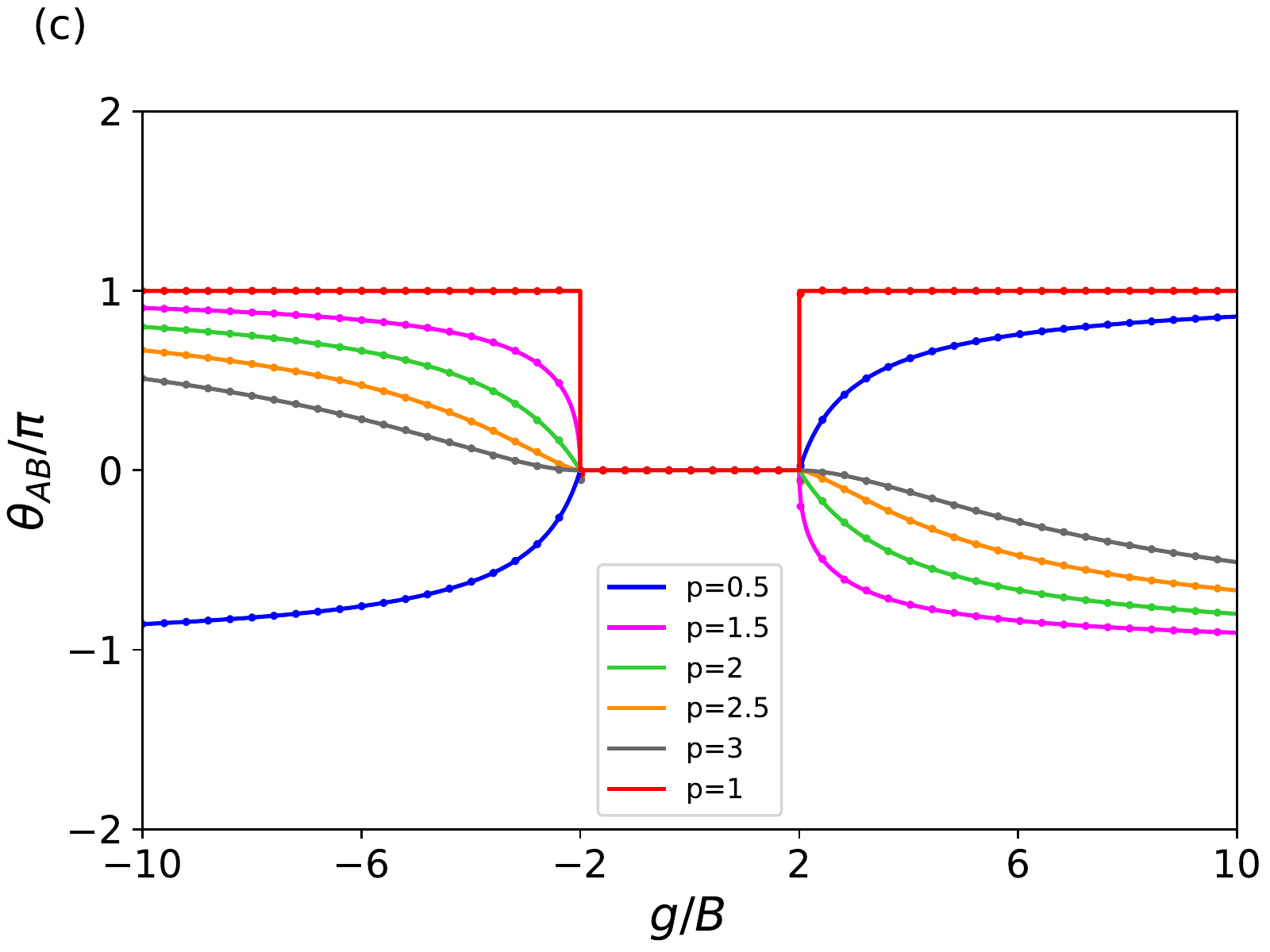}
\end{subfigure}

    \caption{Dynamical, Berry and the nonlinear AB phase plotted against nonlinearity strength $g$ for different values of power law nonlinearity parameter $p$. The solid lines are theoretical values, and the dots are numerical verification.   Only for Kerr nonlinearity $p=1$, the AB phase has a quantized jump of $\pi$ at the critical value $g=\pm 2B$ {and stays at $\pi$ for $|g|> 2B$.} }
    \label{fig3Phases}
\end{figure*}

As detailed in Appendix \ref{appEig}, For $|g|>2B$, a nonlinear Dirac cone is located at the origin. For $|g|<2B$, the only possible solutions to Eq.~(\ref{eqnFullf}) are $x=\pm 1$ and there is no Dirac cone. For $g\in(0,2B)$, we can hence assign $x_{0}=-1$, and for $g\in(-2B,0)$, we may assign $x_{0}=1$. With this convention, it is clear to see that $\theta_{B}$ is constantly $0$ (mod $2\pi$) for $g\in(-2B,2B)$.  The Berry phase $\theta_{B}$ becomes nonzero and changes continuously for $|g|>2B$.   For each $p$, as we continuously tune $g$, $x_{0}$ can be easily solved numerically using Eq.~(\ref{eqngx}), thus obtaining the theoretical values of the leading terms of the dynamical phase, Berry phase and AB phase around the origin. We also numerically solve the evolution using the Schr\"{o}dinger equation Eq.~(\ref{eqnEvolution}) along the two paths, and compute the dynamical phase, AB phase and Berry phase using numerical solutions of the evolution. The evolution is computed using an operator-splitting algorithm. The results are presented in Fig.~\ref{fig3Phases}. 

In each plot, solid lines are theoretical values, while dots on the solid lines are computed from numerical evolutions. In Fig.~\ref{fig3Phases}(a), for any $p=0.5,1,1.5,2,2.5,3$, the dynamical phase around the origin is $0$ for $g\in(-2B,2B)$. At the critical value $g=\pm 2B$ where the Dirac cone appears, for $p=0.5$, the Dirac cone changes continuously with respect to $g$. For $p=1$, there is a quantized jump of $\pm \pi$ at $g=\pm 2B$. For $p=1.5,2,2.5,3$, there is a quantized jump of $\pm 2\pi$ at the critical value $g=\pm 2B$ (so this is equivalent to no change).  In Fig.~\ref{fig3Phases}(b), the Berry phase (modulo $2\pi$) is identically $0$ for $g\in(-2B,2B)$, and changes continuously with respect to $g$. In Fig~\ref{fig3Phases}(c),
the AB phase (modulo $2\pi$) is the sum of the dynamical phase in Fig.~\ref{fig3Phases}(a) and the Berry phase in Fig.~\ref{fig3Phases}(b). Only for $p=1$, the AB phase has a quantized jump of $\pi$ at the critical value $g=\pm 2B$ and {stays at $\pi$ for $|g| > 2B$}, as discovered by Ref.~\cite{nonlinearDiracCones}. For all other values of $p$, the AB phase changes continuously with respect to $g$. The special behavior of $p=1$ is because of the fact that $p=1$ is a critical value for the limit $\lim_{g\to \pm 2B^{\pm}}\delta\theta_{D}$, as will be explained in the next section.

\section{Mechanism of the jump of AB phase at $g=\pm 2B$ for Kerr nonlinearity}
For $p>1$, we can factor out a factor $(1-x_0^2)$ from $\Delta_0$ which cancels the same factor in the numerator of $\delta\theta_{D}$, 
\begin{align}
    \delta\theta_{D} (p>1) &\sim - \pi \frac{(1+x_0)^p - (1-x_0)^p}{(1+x_0)^{p-1} + (1-x_0)^{p-1}} ,
\end{align}
which equals $\mp 2\pi$ or equivalently zero since $x_0=\pm1$, for $|g|=2B$ or when the Dirac cone starts to appear.

likewise, for $p=1$, we have
\begin{align}
    \delta\theta_{D} (p=1) &\sim -x_0 \pi,
    \label{eqndetlaThetaD}
\end{align}
which equals $\mp \pi$ since $x_0=\pm 1$, for  $|g|=2B$. 

Finally, for $0<p<1$, 
\begin{align}
    \delta\theta_{D} (0<p<1) &\sim - \pi (1-x_0^2)^{1-p}\frac{(1+x_0)^p - (1-x_0)^p}{(1+x_0)^{1-p} + (1-x_0)^{1-p}},
\end{align}
which vanishes for $x_0=\pm1$, for  $|g|=2B$. 

Calculations above make it clear that the nonlinear AB phase associated with Kerr nonlinearity ($p=1$) is most special as the extra nonlinearity-induced correction to dynamical phase experiences a $\pi$ jump when the Dirac cone appears. What is intriguing for Kerr nonlinearity is that the nonlinear AB phase stays quantized at $\pi$ for $|g|>2B$, as $\theta_{B}$ and $\delta\theta_{D}$ happen to be complementary to each other, as shown in Eqs.~(\ref{eqnThetaBLeading}) and (\ref{eqndetlaThetaD}).  For all other forms of power-law nonlinearity, there is no such jump,  $\pi$-quantization is thus absent, and consequently, the nonlinear AB phase  only changes continuously with respect to $g$.  This finally explains why in Fig.~\ref{fig3Phases} only the nonlinear AB phase for Kerr nonlinearity ($p=1$) displays a quantization plateau for $|g|>2B$. 

\section{Conclusion}
In this paper, we analytically and computationally examined the so-called nonlinear AB phase around  Dirac cones induced by power-law nonlinearity added to the QWZ model often used for studies of topological band structures.  With our analytical results, we are able to explain why the nonlinear AB phase has a quantized jump of $\pi$ when Dirac cone starts to appear or disappear, for and only for Kerr nonlinearity. In the context of nonlinear AB phase that can be in principle measured in experiments, Kerr nonlinearity is thus identified as a critical form of nonlinearity.

\begin{acknowledgments}
J.G. is grateful to Prof. Giulio Casati for his many years of guidance, interaction and collaboration.   J.G. would also like to thank Prof. Chushun Tian for very useful discussions. 
The computational work for this article was fully performed on resources of the National Supercomputing Centre, Singapore (https://www.nscc.sg).
\end{acknowledgments}

%\newpage

%\nocite{*}
%\newpage

%\newpage
\appendix

\section{Eigenvalue problem}
\label{appEig}
The instantaneous eigenenergy and eigenstate satisfy the Schr\"odinger equation,
\begin{align}
    \begin{bmatrix}
        \beta(k_{1},k_{2})+g|\psi_{1}|^{2p} & \gamma(k_{1},k_{2})\\
        \gamma^{*}(k_{1},k_{2}) &-\beta(k_{1},k_{2})+g|\psi_{2}|^{2p}
    \end{bmatrix}
    \begin{bmatrix}
        \psi_{1}\\
        \psi_{2}
    \end{bmatrix}
    =E\begin{bmatrix}
        \psi_{1}\\
        \psi_{2}
    \end{bmatrix}.
\end{align}
%\begin{align}
%    \beta&=B(-1+\cos k_{1}+\cos k_{2})\label{eqnBeta}\\
%    \gamma&=J_{1}\sin k_{1}-iJ_{2}\sin k_{2}\label{eqnGamma}
%\end{align}
In terms of the two components of the state, we have
%\begin{align}
%    \beta \psi_{1}+g|\psi_{1}|^{2p}\psi_{1}+\gamma\psi_{2}&=E\psi_{1}, \label{eqnEig1}\\
%    \gamma^{*}\psi_{1}-\beta\psi_{2}+g|\psi_{2}|^{2p}\psi_{2}&=E\psi_{2}. \label{eqnEig2}
%\end{align}
\begin{align}
    \gamma\psi_{2}&=(E-\beta-g|\psi_{1}|^{2p})\psi_{1},  \label{eqnEig1}\\
    \gamma^{*}\psi_{1}&=(E+\beta-g|\psi_{2}|^{2p})\psi_{2}. \label{eqnEig2}
\end{align}
Plugging the expressions in Eq.~(\ref{eqn:y20}), we see that
\begin{align}
  \gamma e^{i\varphi} &= \left[E -\beta - g \left(\frac{1+x}{2}\right)^p\right] \sqrt{\frac{1+x}{1-x}}.
\end{align}
Since the right hand side of the above equation is real, we recognize that the phase variable $\varphi$ is simply the opposite of the phase of $\gamma$,
\begin{align}
    \varphi = - \textrm{arg} (\gamma).
\end{align}
Recall that $\gamma = B (\sin k_1 - i \sin k_2)$ in our choice, this means that $\varphi$ is the same angle illustrated in Fig.~\ref{figDynPath}  for sufficiently small $|k_{1}|$ and $|k_{2}|$. 

Multiplying $\psi_{1}^{*}$ on both sides of Eq.~(\ref{eqnEig1}), multiplying $\psi_{2}^{*}$ on Eq.~(\ref{eqnEig2}) and taking complex conjugate, then subtracting the two equations, one ontains
\begin{align}
    \beta+g(|\psi_{1}|^{2p+2}-|\psi_{2}|^{2p+2})&=E(|\psi_{1}|^{2}-|\psi_{2}|^{2}).\label{eqnEig3}
\end{align}
In terms of the parameter $x$ defined in Eq.~(\ref{eqn:y20}), we get the instantaneous eigenenergy as in Eq.~(\ref{eqnEInxFull}).

One can then multiply the two equations in (\ref{eqnEig1}) and (\ref{eqnEig2}) together. Eliminating the common factor $\psi_{1}\psi_{2}$, we arrive at 
\begin{align}
    |\gamma|^{2}&=E^{2}-gE(|\psi_{1}|^{2p}+|\psi_{2}|^{2p})-\beta^{2}-\beta g(|\psi_{1}|^{2p}-|\psi_{2}|^{2p})
    %\nonumber\\ {}&
    +g^{2}|\psi_{1}|^{2p}|\psi_{2}|^{2p}.
\end{align}
Further using Eq.~(\ref{eqn:y20}), we obtain the equation satisfied by the variable $x$ in Eq.~(\ref{eqnFullf}).

Apparently, if the Dirac cone exists, at the Dirac point $k_{1}=k_{2}=0$, the energy is doubly degenerate. As a result, $x$ is also doubly degenerate. Namely, it must satisfy Eq.~(\ref{eqngx}) with $\beta(0,0)=B$. That is
\begin{align}
    2^{p+1}B+ g\left[(1+x)^{p}-(1-x)^{p}\right]&=0. \label{eqnxSolution}
\end{align}
Denote the solution of the above equation as $x_{0}$, i.e.,
\begin{align}
    \left(\frac{1+x_{0}}{2}\right)^{p}-\left(\frac{1-x_{0}}{2}\right)^{p}&=-\frac{2B}{g}. \label{eqnx0}
\end{align}
Note that the left hand side of the above equation is a monotonically increasing function of $x_0$ as $x_{0}\in[-1,1]$, with a minimum of $-1$ and a maximum of $+1$. Therefore,
\begin{align}
    -1\leq -\frac{2B}{g}\leq 1.
\end{align}
This means that
\begin{align}
    |g|\geq 2B. \label{eqnConditiong}
\end{align}
This is the necessary condition for a Dirac cone to exist.

It is also of interest to use the perturbation theory to solve the eigenenergies near the Dirac cone. For sufficiently small $|k_{1}|$ and $|k_{2}|$, we let
\begin{align}
    x&\sim x_{0}+ \chi, \label{eqnxExpansion}\\
    \beta&\sim B + \rho, \label{eqnbetaExpansion}\\
    |\gamma|^{2}& \sim 0 + \eta,\label{eqngamma2Expansion}
\end{align}
where $\chi$ is at least in the first order in $k_{1}$ and $k_{2}$, and $\rho$ and $\eta$ are at least in the second order in $k_{1}$ and $k_{2}$. Plugging Eq.~(\ref{eqnxExpansion}), Eq.~(\ref{eqnbetaExpansion}) and Eq.~(\ref{eqngamma2Expansion}) into Eq.~(\ref{eqnFullf}), we have
\begin{align}
     \frac{g^2 p^2}{2^{2p+2}} \frac{1-x_{0}^{2}}{x_0^2} \left[(1+x_{0})^{p-1}+(1-x_{0})^{p-1}\right]^2\chi^{2} 
     &\sim \eta .
\end{align}
To this order, we get the correction to the parameter $x$,
\begin{align}
    \chi &\sim \pm \frac{2^{p+1}}{gp} \frac{x_{0}}{(1+x_{0})^{p-1}+(1-x_{0})^{p-1}} \frac{\sqrt{J_{1}^{2}k_{1}^{2}+J_{2}^{2}k_{2}^{2}}}{ \sqrt{1-x_{0}^{2}}}.
\end{align}
Plugging this into Eq.~(\ref{eqnEInxFull}), we find the expression for the eigenenergy, 
\begin{align}
    E&\sim E_{0} \left(1+ p \frac{\chi}{x_0}\right),
\end{align}
where the nonperturbed eigenenergy is
\begin{align}
    E_{0}&= \frac{g}{2} \left[\left(\frac{1+x_0}{2}\right)^p + \left(\frac{1-x_0}{2}\right)^p \right] .
\end{align}

\begin{figure*}[htpb!]
\centering
\begin{subfigure}{1\columnwidth}
\centering
\includegraphics[width=0.45\textwidth]{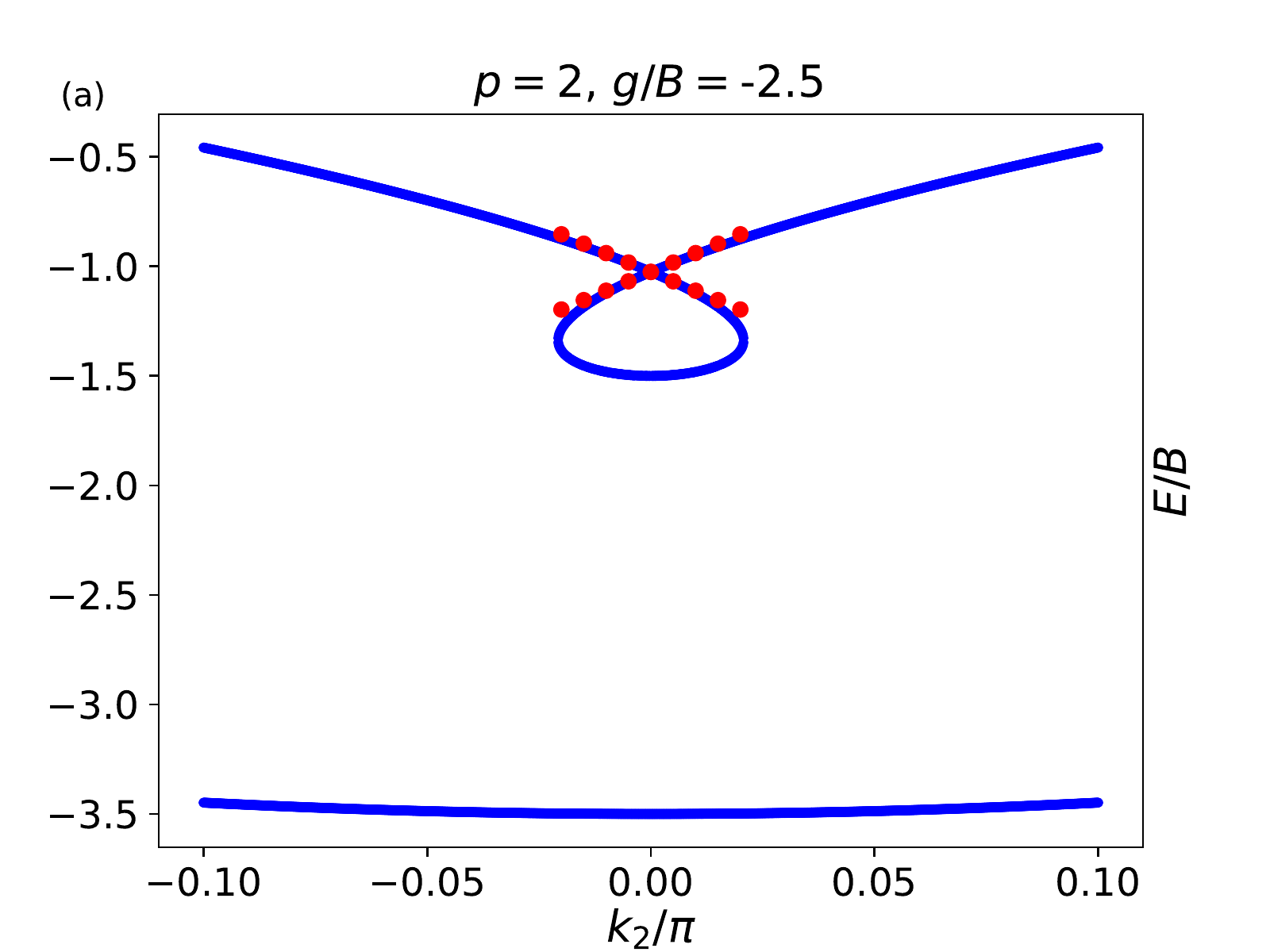}
%\end{subfigure}\hfill
%\begin{subfigure}{1\columnwidth}
\centering
\includegraphics[width=0.45\textwidth]{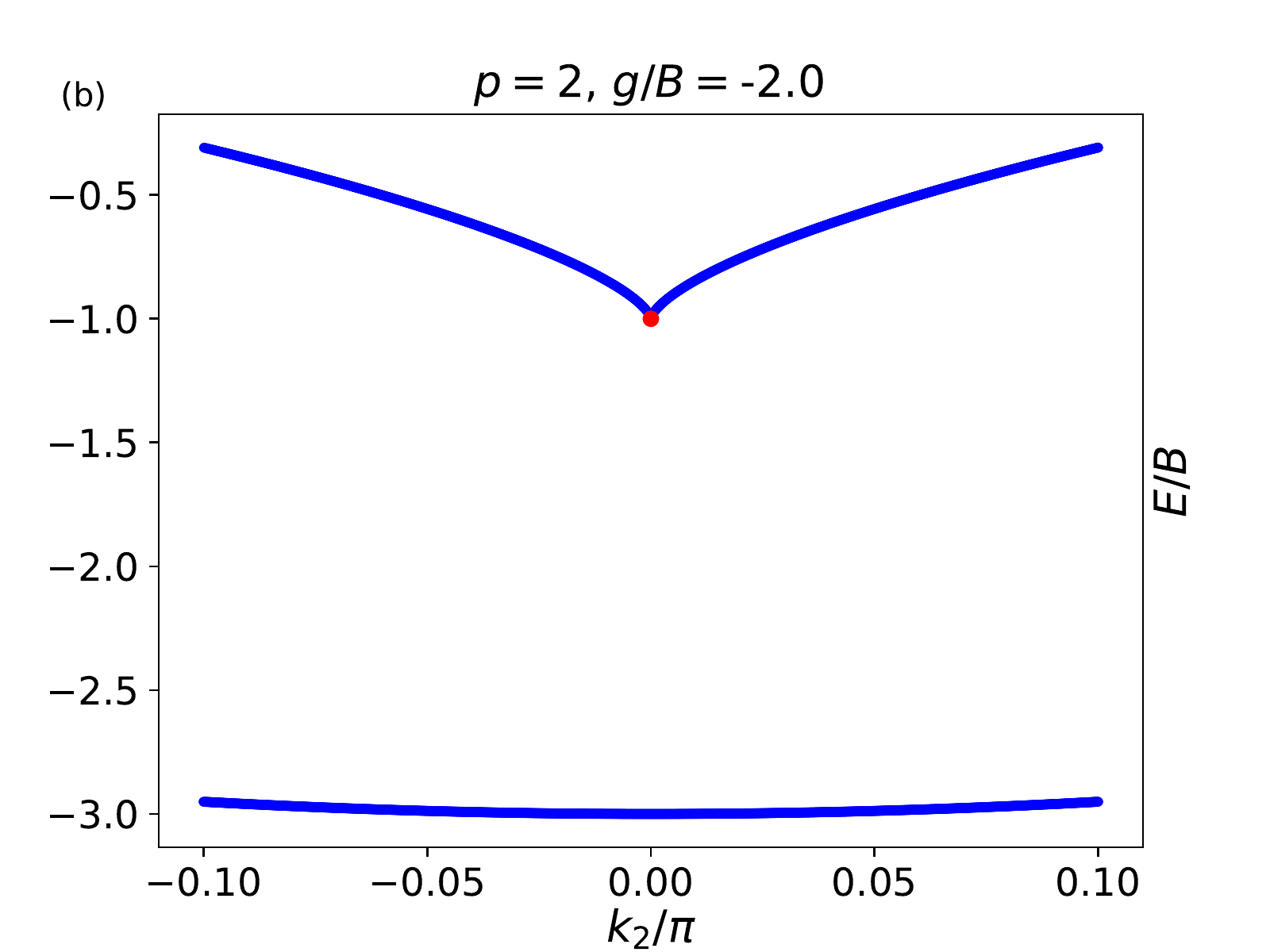}
\end{subfigure}

\medskip

\begin{subfigure}{1\columnwidth}
\centering
\includegraphics[width=0.45\textwidth]{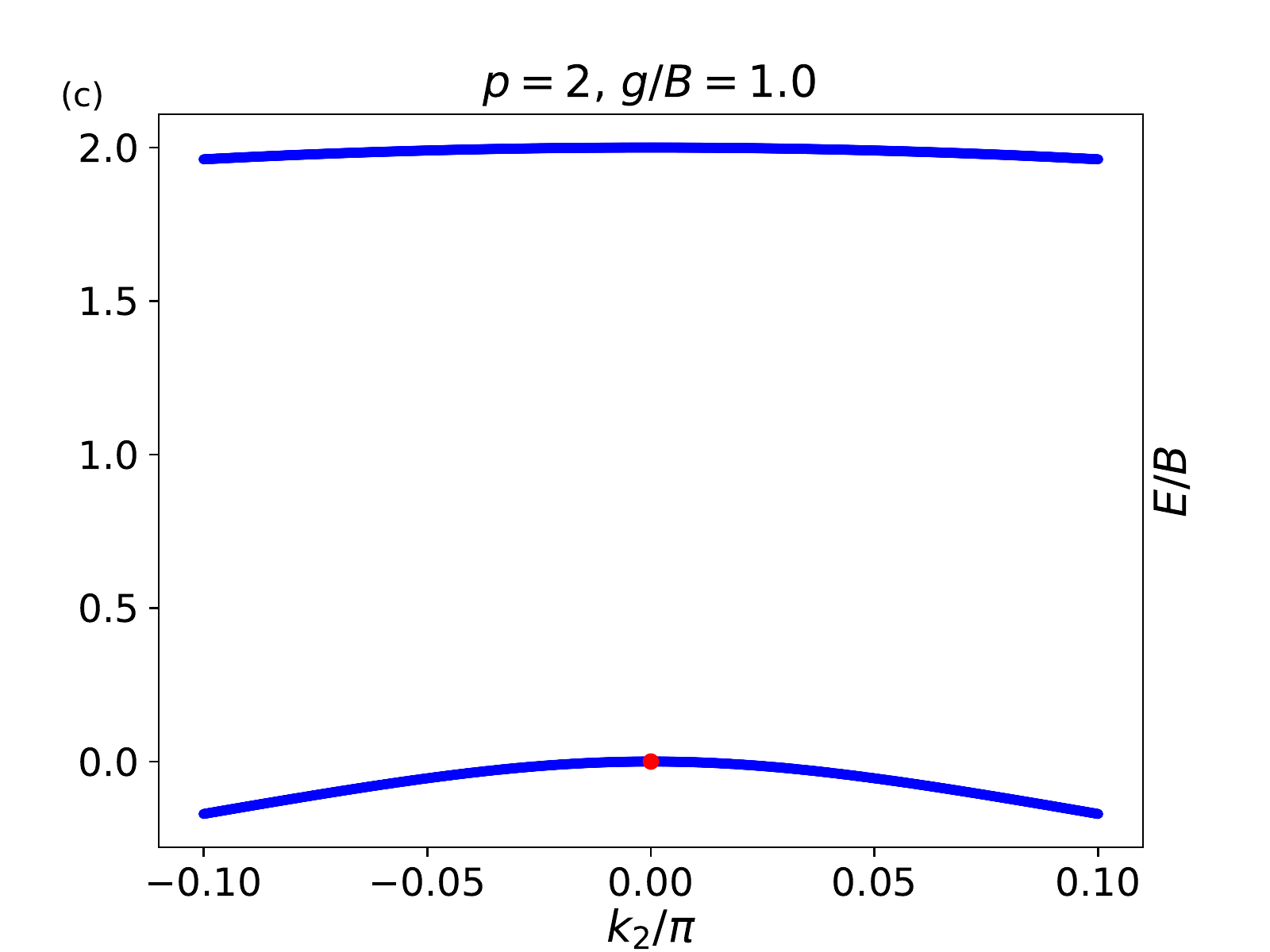}
%\end{subfigure}\hfill
%\begin{subfigure}{1\columnwidth}
\centering
\includegraphics[width=0.45\textwidth]{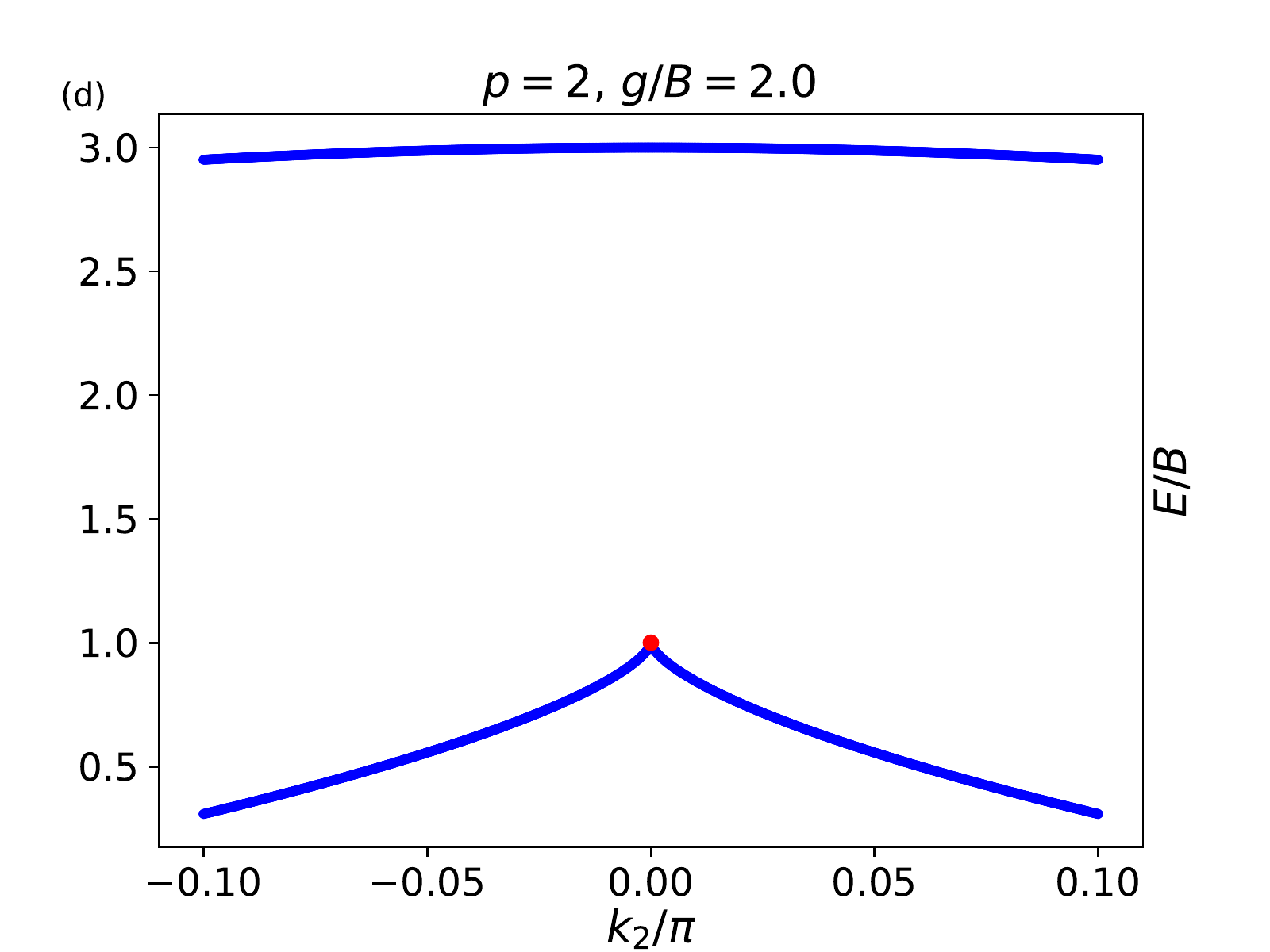}
\end{subfigure}

\medskip

\begin{subfigure}{1\columnwidth}
\centering
\includegraphics[width=0.45\textwidth]{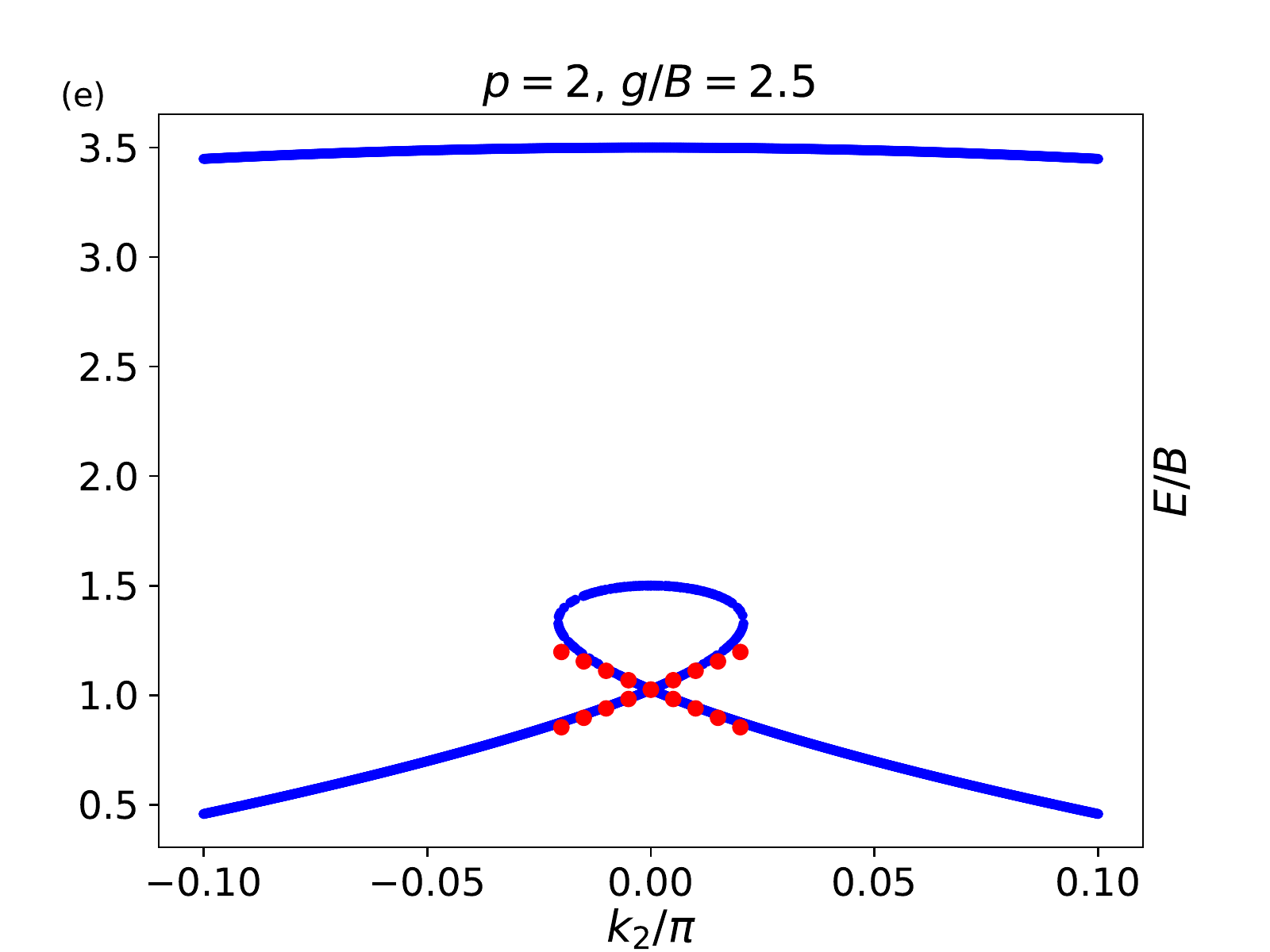}
\end{subfigure}

\caption{Numerical and perturbative solutions of $E$ for $p=2$ with $g/B=-2.5,-2,1,2,2.5$, $B=2$ along section $k_{1}=0$. In these plots, the red dots are perturbative eigenenergies near the Dirac cone and the blue lines are numerical solutions.}
\label{figSpectrump2}

\end{figure*}

We can see clearly from the expansion of $E$ that there is a Dirac cone structure at the origin, provided $|x_{0}|<1$, which corresponds to $|g|>2B$. For $|g|<2B$, the system contains two smooth energy bands. At the critical value $g=2B$ ($g=-2B$), a kink will develop on the lower (upper) band at $k_{1}=k_{2}=0$. Once $g>2B$ ($g<-2B$), a 2D self-intersection structure, i.e., a nonlinear Dirac cone, will appear from the lower (upper) band, whose vertex is at $k_{1}=k_{2}=0$. This is true for any $p>0$. We show 5 plots with different values of nonlinearity in Fig.~\ref{figSpectrump2}, along section $k_{1}=0$ and with $-0.1\pi \leq k_{2}\leq 0.1\pi$. In each plot, red dots are perturbative eigenenergies around the Dirac point (or at the origin for $|g|\leq 2B$), while blue lines are numerical solutions by solving Eq.~(\ref{eqnFullf}) exactly. We can see that the perturbative solutions perfectly match the numerical solutions for sufficiently small $|k_{2}|$.
%\begin{figure}[hbtp!]
	%\begin{centering}
		%\includegraphics[scale=0.3]{pics/spectrump2/g-2.5B.png}\includegraphics[scale=0.3]{pics/spectrump2/g-2.0B.png}\includegraphics[scale=0.3]{pics/spectrump2/g1.0B.png}
	%\par\end{centering}
	%\begin{centering}
		%\includegraphics[scale=0.3]{pics/spectrump2/g2.0B.png}
	%\includegraphics[scale=0.3]{pics/spectrump2/g2.5B.png}
		%\par\end{centering}
	%\caption{Numerical and perturbative solutions of $E$ for $p=2$ with $g/B=-2.5,-2,1,2,2.5$, $B=2$ along section $k_{1}=0$}\label{figSpectrump2}
%\end{figure}

%\section[\appendixname~\thesection]{Dynamics around the Dirac cone}

\section{Dynamics around the Dirac cone}
\label{appDynamics}
We solve the following Schr\"{o}dinger equation perturbatively in terms of adiabatic parameter $\varepsilon$ \cite{nonlinearDiracCones},
\begin{align}
    i\partial_{t}\begin{bmatrix}
        \Psi_{1}\\
        \Psi_{2}
    \end{bmatrix}=\begin{bmatrix}
        \beta & \gamma\\
        \gamma^{*} &-\beta
    \end{bmatrix}
    \begin{bmatrix}
        \Psi_{1}\\
        \Psi_{2}
    \end{bmatrix}+g\begin{bmatrix}
        |\Psi_{1}|^{2p}\Psi_{1}\\
        |\Psi_{2}|^{2p}\Psi_{2}
    \end{bmatrix}.
    \label{eqn:Schro}
\end{align}
During the adiabatic following process, the quasimomenta are tuned adiabatically, 
\begin{align}
    k_{1}&=k_{1}(\varepsilon t), \qquad 
    k_{2}=k_{2}(\varepsilon t),
\end{align}
with $0<\varepsilon\ll 1$. Let
\begin{align}
    \ket{\Psi} &= 
    \begin{bmatrix}
        \Psi_1\\
        \Psi_2
    \end{bmatrix} = e^{i\theta}(\ket{\psi} + \varepsilon \ket{\phi}) 
    = e^{i\theta} 
    \begin{bmatrix}
        \psi_1\\
        \psi_2
    \end{bmatrix}
    + \varepsilon e^{i\theta}
    \begin{bmatrix}
        \phi_1\\
        \phi_2
    \end{bmatrix}, \label{eqnPsia}
\end{align}
with
\begin{align}
    \dot{\theta}&\sim -E + \varepsilon\alpha, 
    \label{eqn:theta}
\end{align}
where $\psi_{a}$ are the solutions to the eigenvalue problem in Eq.~(\ref{eqnEigOriginal}), and $\varepsilon \phi_{a}$ are the first order corrections. As we shall see, the dynamical phase comes from both $E$ and $\varepsilon\alpha$, and geometric phase comes from only $\varepsilon\alpha$. 

The solution to the adiabatic process deviates from the instantaneous eigensolution by the order of $\varepsilon$, multiplied by a phase term. Note that the increment of phase $\theta$ over a small duration of time $dt$ may deviate from the contribution of dynamical phase by $O(\varepsilon)$, but the accumulation of the $O(\varepsilon)$ term over the total time $O(\varepsilon^{-1})$ has a contribution of $O(1)$. 

The Hamiltonian can be expand in the power of $\varepsilon$ accordingly,
\begin{align}
    \widehat{H}(\Psi) &\sim \widehat{H}(\psi) + \varepsilon \hat{h}(\psi,\phi),
\end{align}
where $\widehat{H}(\psi)$ is given in Eq.~(\ref{eqnHamiltonian}) and $\hat{h}$ depends on both $\psi$ and $\phi$ with a diagonal form,
\begin{widetext}
\begin{align}
    \hat{h}(\psi,\phi) & = gp \begin{bmatrix}
        |\psi_1|^{2p-2} (\psi_1^* \phi_1 + \psi_1 \phi_1^*) & 0\\
        0 & |\psi_2|^{2p-2} (\psi_2^* \phi_2 + \psi_2 \phi_2^*)
    \end{bmatrix}.
\end{align}
\end{widetext}
To compute $\hat{h}$, we need to expand $|\Psi|^{2p}$. To the first order in $\varepsilon$, we have
\begin{align}
    |\psi_a + \varepsilon \phi_a|^{2p} &\sim |\psi_a|^{2p} \left[1 + 2\varepsilon p \operatorname{Re} \left(\frac{\phi_a}{\psi_a}\right)\right] \nonumber\\
    &= |\psi_a|^{2p} + \varepsilon p|\psi_a|^{2p-2} (\psi_a^* \phi_a + \psi_a \phi_a^*).
    \label{eqn:HHh}
\end{align}
Plugging Eqs.~(\ref{eqnPsia}), (\ref{eqn:theta}) and (\ref{eqn:HHh}) into the time-dependent Schr\"{o}dinger equation in Eq.~(\ref{eqn:Schro}), up to the first order in $\varepsilon$, we obtain

\begin{align}
    %{}&
    E \ket{\psi} -\varepsilon \alpha \ket{\psi} + \varepsilon E \ket{\phi} + i\ket{\dot{\psi}} + i \varepsilon\ket{\dot{\phi}} 
    %\nonumber\\ &{}
    &\sim \widehat{H}(\psi) \ket{\psi} + \varepsilon \hat{h}\ket{\psi} + \varepsilon \widehat{H}(\psi) \ket{\phi}.
    \label{eqn:bigSchro}
\end{align}
Note that the time derivative brings a factor of $\varepsilon$ because we are in the adiabatic regime, thus the term $\varepsilon\ket{\dot{\phi}}$ is actually in the order of $\varepsilon^2$ and it can be discarded. Apply the instantaneous eigenvalue equation in Eq.~(\ref{eqnEigOriginal}), we get the equation for $\ket{\phi}$,
\begin{align}
    \left[E-\widehat{H}(\psi)\right] \ket{\phi} &\sim \alpha \ket{\psi} - \frac{i}{\varepsilon} \ket{\dot{\psi}} + \hat{h}\ket{\psi}.
    \label{eqn:ketphi}
\end{align}
Multiply $\bra{\psi}$ from the left to Eq.~(\ref{eqn:ketphi}), we get
\begin{align}
    \varepsilon\alpha &= i  \braket{\psi|\dot{\psi}} - \varepsilon \bra{\psi}\hat{h}\ket{\psi}.
\end{align}
After a lengthy calculation, we find the solution to the correction of wavefunction as
\begin{align}
    \varepsilon \phi_1 & = -\frac{x(1-x)\sqrt{1+x}}{4\sqrt{2}} \frac{\dot{\varphi}}{\Delta}  - i \frac{x}{4\sqrt{2(1+x)}} \frac{\dot{x}}{\Delta'}, \\
     \varepsilon \phi_2 & = \frac{x(1+x)\sqrt{1-x}}{4\sqrt{2}} \frac{\dot{\varphi}}{\Delta} e^{i\varphi} + i \frac{x}{4\sqrt{2(1-x)}} \frac{\dot{x}}{\Delta'} e^{i\varphi},
\end{align}
where we introduce two quantities, $\Delta$ in Eq.~(\ref{eqn:Delta}) and $\Delta'$ is defined as
\begin{align}
    \Delta' &:= \beta+\frac{g}{2}\left[\left(\frac{1+x}{2}\right)^p-\left(\frac{1-x}{2}\right)^p\right].
\end{align}
It turns out that the $\dot{x}/\Delta'$ terms do not contribute to $\hat{h}$,
\begin{align}
    \hat{h} = gp\frac{x(1-x^2)}{4\Delta}\frac{\dot{\varphi}}{\varepsilon}
    \begin{bmatrix}
        -\left(\frac{1+x}{2}\right)^{p-1} & 0 \\
        0 & \left(\frac{1-x}{2}\right)^{p-1}
    \end{bmatrix}.
\end{align}
Putting all together, the change rate of the overall phase is given in Eq.~(\ref{eqn:dynphase}).

%\newpage

%23
%\bibliography{ref}


%
\end{document}